\documentclass[prb,showpacs]{revtex4}

\usepackage{graphicx}
\newcommand{\gaf}{\mbox{\boldmath$\gamma$}}
\begin{document}

\title{Eliashberg equations derived from the functional renormalization group}
\author{Carsten Honerkamp$^{1}$ and Manfred Salmhofer$^{2}$}
\affiliation{$^1$ Max--Planck--Institut f\" ur Festk\" orperforschung, 
Heisenbergstrasse 1, D-70569 Stuttgart, Germany
\\ $^2$ Theoretische Physik, Universit\"{a}t Leipzig, 
Augustusplatz 10, D-04109 Leipzig, Germany \\
\small
}
\date{\today}

\begin{abstract}
We describe how the fermionic functional renormalization group (fRG) flow 
of a Cooper+forward scattering problem can be continued into the superconducting state. 
This allows us to reproduce from the fRG flow the fundamental equations of the 
Eliashberg theory for superconductivity at all temperatures including the symmetry-broken 
phase. 
We discuss possible extensions of this approach like the inclusion of vertex corrections.
\end{abstract}


\maketitle
\section{Introduction}
A considerable part of the current understanding of condensed
matter is based on mean-field theories. For superconductivity, the
Bardeen-Cooper-Schrieffer (BCS) theory or its counterpart with
a realistic phonon spectrum and Coulomb repulsion, 
the Eliashberg theory\cite{eliashberg,scalapino,mahan,carbotte}, 
is a powerful framework
that allows one to calculate many properties of superconductors
such as transition temperatures and excitation gaps even
quantitatively\cite{reinert} from a clearly defined starting point. 
Yet in many modern
materials these standard pictures seem to be challenged. In each
case the question arises if the discrepancies arise due to
completely new physics or if the established methods fail only 
partially and can be adapted. A mean-field approach can be
expected to be good if  the fluctuations about it are not too strong and 
if the correlations entailed by the chosen
mean-field are far stronger than other types of correlations. Also this second 
requirement is fulfilled less clearly in many low-dimensional
or strongly correlated materials, where one often sees a
competition between various ordering tendencies at low
temperatures. Hence, in order to obtain an understanding of the
situation it is necessary to find a theoretical description which
can include all the important channels. Renormalization
group (RG) approaches have the ability to treat a competition between
various types of fluctuations in great detail. They allow for an
unbiased detection and qualitative comparison of Fermi liquid
instabilities. This has been demonstrated in many works on
one-dimensional systems\cite{solyom,lin} and the two-dimensional
Hubbard model\cite{zanchi,halboth,honerkamp,tflow,tsai1,kataninka}. 

Recently a functional renormalization method has been applied to electrons
coupled to phonons\cite{tsai}. In this work the authors describe
how Eliashberg theory can be reproduced for temperatures at and
just above the superconducting transition, but not below it. 
Although this work develops an essentially correct and useful physical picture, it
highlights a general drawback of the approximate methods\cite{zanchi,halboth,honerkamp,tflow,tsai1,tsai,kataninka} used until now. In these approaches the renormalization group flow
cannot be extended into the symmetry-broken phase. A notable
exception are theories where the four-fermion interactions are
bosonized at some scale\cite{bick}, but then one has to work hard to remove the 
bias introduced by that into the flow. In addition, for many problems the type of
Hubbard-Stratonovitch decoupling is not obvious from the outset. 
Hence, although several routes seem worth pursuing, a continuation of the 
flow in the unbiased fermionic picture is desirable.  
In the language of flowing coupling constants in a purely
fermionic RG, a flow to strong coupling occurs when the
interactions seem to diverge at a finite energy scale and
consequently the perturbative flow has to be stopped. In many
cases the flow to strong coupling takes place only in a well defined
channel of the fermionic interactions and only one specific
susceptibility becomes large. This can then be interpreted as an
indication for spontaneous symmetry breaking in the corresponding
channel. The energy or temperature scale where this happens can be
taken as an upper estimate for the ordering temperature. 

In a previous work\cite{gapflow} we have developed a method that allows us
to continue the fermionic RG flow into the symmetry broken phase. 
The idea is to include a small symmetry breaking field in the initial conditions of 
the flow. This small offdiagonal selfenergy, e.g. a small superconducting gap, 
grows at the scale where the flow to strong coupling takes off but -- if one uses a 
reorganization of the flow equations proposed by Katanin\cite{katanin} -- prevents 
a divergence of the interactions at nonzero RG scale. 
This allows us to integrate out all modes down 
to zero scale and the interactions and the offdiagonal selfenergy saturate at finite 
values. The Goldstone boson related to the broken symmetry has a small mass gap 
due to the initial symmetry-breaking field. This can be sent to zero afterwards. 
In Ref.
\onlinecite{gapflow} we only considered the reduced BCS model with a
static attraction and showed how the exact gap value of the BCS theory is recovered. 
Here we generalize the approach to include dynamical phonons and the forward scattering 
channel. This yields an additional equation for the normal selfenergy. 
Our extended RG scheme allows us to reproduce the
Eliashberg equations not only for the symmetric phase above the critical temperature as 
in Ref. \onlinecite{tsai} but at all temperatures. 
Furthermore our treatment makes  clear which approximations are used and what one can 
do to go beyond Eliashberg theory.

This paper is organized as follows. In Section \ref{sec2} we briefly introduce the 
model and the Eliashberg equations. In Sec. \ref{sec3} we describe the general 
fermionic functional RG formalism. In Sec. \ref{sec4} we apply an approximate 
version of this formalism to the Eliashberg problem. In Sec. \ref{sec5} we conclude 
with a discussion of future extensions of the approach.

\section{The Model and Eliashberg theory} \label{sec2}
\begin{figure}
\begin{center}
\includegraphics[width=.5\textwidth]{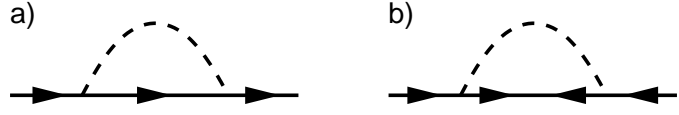}
\end{center}
\caption{Diagrammatic expression for the self-consistent Eliashberg equations. 
The internal solid lines are full normal (a)) and anomalous (b)) propagators. 
The dashed line denotes the effective interaction. }
\label{EliashFig}
\end{figure}

The model we study is the basically same as in Ref. \onlinecite{tsai}. We
consider spin-$1/2$ electrons with a dispersion $\epsilon
(\vec{k})$. In order to keep the formalism simple we restrict
ourselves to a spherically symmetric system with a 
smooth and finite density of states in the energy window of interest around 
the Fermi level. The electron--electron interaction is given by a static and
spin-rotationally invariant interaction $V_c = u(k_1,k_2,k_3,k_4)$.
 We have written $k_i$
for the Matsubara frequency and wave-vector of electron $i$. In
our notation, $k_1$ and $k_2$ belong to the two incoming electrons
and the spin component of $k_1$ and the first outgoing particle
$k_3$ is the same. 
$V_c$ can be envisaged as a screened Coulomb interaction.
Its precise form is not needed in this paper.
Next we add to $V_c$ a phonon-mediated interaction
\begin{equation}
V_{\mathrm{ph}} (k_1,k_2,k_3,k_4) = - g (k_1,k_3) g(k_2,k_4) D(k_1-k_3)
\end{equation}
which arises due to the exchange of phonons with
propagator 
\begin{equation} 
D(q_0,\vec{q}) = \int_0^\infty d \omega \, B(\omega,\vec{q}) 
\left[ \frac{1}{iq_0 - \omega} - \frac{1}{iq_0 + \omega} \right] . \end{equation}
$g(k_1,k_3)$ is the coupling to the fermions. It is well known
that the exchange of phonons can induce superconductivity. In
this case the selfenergy of the electrons acquires a nonzero
off-diagonal part $\Delta (\omega )$ in addition to a
quasiparticle renormalization factor $Z(\omega)$ in the diagonal
selfenergy. The Eliashberg theory aims at calculating $\Delta (\omega )$ 
and  $Z(\omega)$ for a given phonon spectrum $B(\omega,\vec{q})$. 
It is contained in a set of two
self-consistent equations for the normal self-energy $\Sigma (k)$ and the 
anomalous self-energy $\Delta (k)$ which read\cite{mahan} 
(assuming spin-singlet pairing for simple notation)
\begin{eqnarray}
{\Sigma} (k) &=& - \sum_{k'}   V_{\mathrm{eff}} (k,k',k',k) G(k')  \label{E1}
\\
{\Delta} (k ) &=& - \sum_{k'}   V_{\mathrm{eff}} (k,-k,k',-k') F(k') \label{E2}
\end{eqnarray}
$G(k)$ and $F(k)$ are the diagonal and off-diagonal propagators, respectively, and  \begin{equation}
 V_{\mathrm{eff}}(k_1,k_2,k_3,k_4)=u(k_1,k_2,k_3,k_4) + V_{\mathrm{ph}}(k_1,k_2,k_3,k_4) . \end{equation}
In many cases it is a good approximation to assume that the 
effective interaction depends only on the frequency-momentum transfer $k_1-k_3$. 
The two equations (\ref{E1}) and (\ref{E2}) can be visualized in two 
Fock-type self-energy diagrams as shown 
in Fig. \ref{EliashFig}. The $Z$-factor is found from the normal self-energy via 
\begin{equation}
{\Sigma} (k)= {\Sigma}_e(\vec{k}) +i k_0 \left[ 1 - Z(k) \right] \, . 
\end{equation}
The even-frequency part $\Sigma_e(k)$ is typically weakly $k$-dependent\cite{eliashberg}. 
Then it is treated as a renormalization of the chemical potential and is not considered 
further\cite{mahan}. The spectral gap is determined by the retarded gap function 
\begin{equation}
\tilde \Delta (k) = \frac{{\Delta}(k)}{Z(k)} \, . 
\end{equation}
In these equations, phonon vertex corrections for the phonon-electron vertex are 
neglected based on Midgal's theorem. The dressing by the Coulomb interaction is 
absorbed into an effective electron-phonon matrix element.
The Coulomb vertex is taken as a constant in the static limit\cite{scalapino}.
In this approximation, the Hartree selfenergy contribution is ignored. 
The conventional argument is that the phonon-mediated interaction  in this diagram 
does not produce a frequency dependence of the selfenergy and hence no contribution 
to $Z$, and possible frequency dependences from the renormalized Coulomb repulsion 
are argued to be less relevant for the questions of interest. 
However for general situations, this diagram can still cause Fermi surface deformations. 
Hence in problems where the Fermi surface geometry and location are important, 
the Hartree term should be taken into account.

\section{Functional RG} \label{sec3}
\begin{figure}
\begin{center}
\includegraphics[width=.8\textwidth]{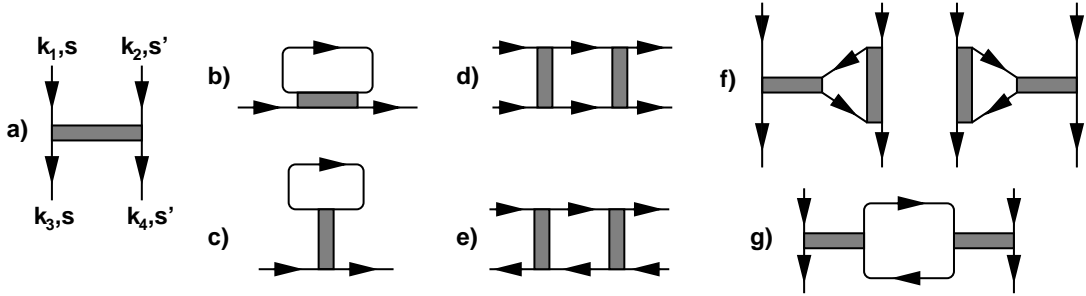}
\end{center}
\caption{a) Coupling function $V_\Lambda (k_1,k_2,k_3)$. b) Fock diagram for the 
selfenergy. c) Hartree diagram for the selfenergy. d) Particle-particle diagram, 
e) crossed particle-hole diagram, f) vertex corrections, 
g) electronic corrections to the phonon propagator or screening. }
\label{5dias}
\end{figure}

The backbone of our approach is the functional renormalization
group technique for 1-particle irreducible (1PI) vertex functions
\cite{wetterich,salmhofer}. It is derived from an exact equation
for the generating functional for the 1PI vertex functions of the
theory when a flow parameter in the quadratic part ${\bf Q}$ 
of the action is changed. Inserting into this equation an expansion of the
generating functional in monomials of the source fields with the
$n$-point vertex functions $\tilde{\gaf}^{(n)} (K_1,K_2, \dots , K_n)$
as coefficients one obtains an infinite hierarchy of flow
equations for the $n$-point vertices. The index
$K=(c,s,\vec{k},i \omega)$ comprises the
wave-vector $\vec{k}$, Matsubara frequency $i \omega$, spin $s$
and the Nambu particle-hole index $c$ which is $c=+$ for a fermionic
field $\bar \psi (\vec{k},i \omega, s)$ and $c=-$ for $\psi
(-\vec{k},-i \omega, -s)$. All bold-face quantities carry these Nambu indices.
We shall treat the translation invariant case. Then
\begin{equation}
{\bf \gaf}^{(n)} (K_1,K_2, \dots , K_n)
=
\delta(c_1k_1 + \ldots + c_n k_n)
\tilde{\gaf}^{(n)} (K_1,K_2, \dots , K_n),
\end{equation} 
${\bf G} (K_1,K_2) = \delta(c_1k_1+c_2k_2) \tilde {\bf G} (K_1,K_2)$,
and similarly for ${\bf S}$. Here we used $K_l = (c_l,s_l, k_l)$ with
$k_l = (\vec{k}_l, i \omega_l)$. 
This hierarchy of equations is, in a
first approximation, truncated after the irreducible 4-point
vertex. Then one is left with one flow equation for the 2-point
vertex, which gives the flow of the self-energy, and one equation
for the flow of the four-point vertex. They read
\begin{equation}
\dot {\bf \Sigma} (K_1,K_2) = - \frac{1}{2} \, \sum_{P_1,P_2}
\gaf^{(4)} (K_1,K_2,P_1,P_2)
{\bf S}( P_1,P_2) 
\label{sigmadot} \end{equation}
\begin{eqnarray}
\dot{\gaf}^{(4)} (K_1,K_2,K_3,K_4) &=& -
 \frac{1}{2} \sum_{P_1,P_2,P_3,P_4}  {\bf L}(P_1,P_2,P_3,P_4) \nonumber \\ 
 && \cdot \left[
 \gaf^{(4)} (K_1,K_2,P_2,P_3) \gaf^{(4)} (P_4,P_1,K_3,K_4) \right. 
 \nonumber\\ && -\, \gaf^{(4)} (K_1,K_3,P_2,P_3) \gaf^{(4)} (P_4,P_1,K_2,K_4)
\nonumber \\ && \left. +\, \gaf^{(4)} (K_1,K_4,P_2,P_3) \gaf^{(4)} (P_4,P_1,K_2,K_3)
 \right]
\label{gamma4dot}
\end{eqnarray}
The dot denotes the derivative $d/d\Lambda$ with respect to the RG scale which we choose as an infrared cutoff $\Lambda$.
${\bf S}(K_1,K_2)$ is the so-called single-scale
propagator\cite{salmhofer}
\begin{equation}
{\bf S}(K_1,K_2) = - \sum_{K,K'} {\bf G}(K_1,K) \dot {\bf Q}(K,K')  {\bf G}(K',K_2) 
\end{equation}
with the full scale-dependent propagator ${\bf G}(K_1,K_2)$.
 ${\bf L}(P_1,P_2,P_3,P_4)$ 
is the scale-derivative of the
product of two full Green´s functions,
\begin{equation}\label{eq:10}
{\bf L}(P_1,P_2,P_3,P_4) = \frac{d}{d\Lambda} \left[ {\bf G}(P_1,P_2){\bf G}(P_3,P_4)
\right] \end{equation}
Equation (\ref{eq:10}) corresponds to the
modified 1PI-RG scheme as introduced by Katanin\cite{katanin} and
discussed thoroughly in Ref.\ \onlinecite{gapflow}. This modification is
essential in order to obtain correct results in the
symmetry-broken phase. The single-scale propagator is related to the scale-derivative 
of the Green's function by
\begin{equation}
\dot {\bf G}(K_1,K_2) = {\bf S}(K_1,K_2) + \sum_{K,K'} {\bf G}(K_1,K) \dot {\bf \Sigma}(K,K')  {\bf G}(K',K_2)
\label{gdots} 
\end{equation}
In the symmetric phase, the spin-rotationally invariant interaction vertex can be 
expressed\cite{salmhofer} by a coupling function $V_\Lambda(k_1,k_2,k_3)$ where the 
spin indices $s$ ($s'$) of the first (second) incoming and the first (second) outgoing 
particles are the same (see Fig. \ref{5dias} a)). That is, 
\begin{equation}
\gamma^{(4)} 
\big(
(+,s,k_1),(+,s',k_2),(-,s',k_3),(-,s,k_4)
\big)
=
\delta(k_1 + k_2 - k_3 -k_4)\;
V_\Lambda(k_1,k_2,k_3),
\end{equation}
so that $V_\Lambda (k_1,k_2,k_3)$ describes the scattering 
$(k_1,s) \to (k_3,s)$ and $(k_2,s') \to (k_4,s')$,
and all other values of $\gamma^{(4)}$ are fixed by the fermionic 
antisymmetry and the invariance under spin rotations and charge 
conjugation.   
With the assumptions of the previous section we can write the initial (bare) coupling 
function as
\begin{equation} 
V_{\Lambda_0}  (k_1,k_2,k_3) = V_{\mathrm{eff}} (k_1,k_2,k_3,k_4) 
= u (k_1,k_2,k_3,k_4) - g(k_1,k_3) g(k_2,k_4) D(k_1-k_3) \, . 
\label{initialint}
\end{equation}
Actually the precise form of the initial interaction is not important,
provided it is not long--range, i.e.\ singular in momentum space,
and in the following we just study a general regular four--point term
$V_{\Lambda_0}  (k_1,k_2,k_3)$. In fact, the interaction corresponding
to an exchange of acoustic phonons can be treated without difficulty 
even though its derivatives in momentum space are unbounded. 
If one limits the considerations to singlet superconducting pairing, 
spin-rotation symmetry holds at all temperatures. Then the parameterization 
of the normal interaction vertex can still be used in the U(1)-broken phase. 
In this phase however, anomalous interaction vertices are generated with an 
unequal number of incoming and outgoing outgoing lines.

\begin{figure}
\begin{center}
\includegraphics[width=.7\textwidth]{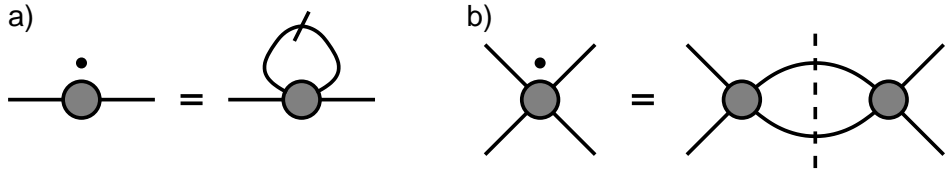}
\end{center}
\caption{RG equations for the two-point a) and the four-point vertex b). The slashed line denotes a single-scale propagator $S_\ell(p)$ while the dashed line symbolizes the scale derivative of the one-loop diagram. The one-loop graph in b) includes particle-particle and particle-hole contributions. With our truncation $\gaf^{(m)}_\ell=0$ for $m\ge 6$ the feedback of the $\gaf^{(6)}_\ell$ on $\gaf^{(4)}_\ell$ is neglected.}
\label{RGeqs}
\end{figure}
Diagrammatically the equations above are shown in Fig.
\ref{RGeqs}. In this form, the lines in the diagrams do not have a
direction yet as the fermionic fields still carry the Nambu index.
Resolving the Nambu index for the U(1)-symmetric normal phase we obtain two 
diagrams for the self-energy, one Hartree and one Fock term, and 5 diagrams 
for the flow of the interaction. These contain particle-particle and particle-hole 
diagrams (see Fig. \ref{5dias}). 

\section{Cooper + Forward scattering model} \label{sec4}
Now we restrict the analysis to a spherical Fermi surface and energy scales
that are low compared to the Fermi energy. Furthermore we concentrate on systems 
which do not violate spin-rotation invariance.
In the low--energy regime the theory simplifies drastically 
because most contributions are subleading and can be dropped. 
This allows us to reproduce the Eliashberg equations. 
Let us consider the flow in the normal phase first.

At scale $\Lambda$ in the flow, we have already integrated out all degrees of
freedom with energies above $\Lambda$. For small $\Lambda$, 
only a thin shell of width $\Lambda \ll E_F$ is left to integrate over. 
The geometry of this thin shell leads to strong kinematical constraints
if the Fermi surface is convex and positively curved \cite{feldman,shankar,chen,metzner}, 
and there are only three classes of two-particle interaction processes 
where all incoming and outgoing momenta are close to the Fermi surface. 
These are the Cooper processes with zero total incoming wavevector 
$\vec{k}_1+\vec{k_2}=0$, the direct forward scattering with $\vec{k}_1=\vec{k}_3$, 
$\vec{k}_2=\vec{k}_4$, and the exchange forward scattering 
$\vec{k}_1=\vec{k}_4$, $\vec{k}_2=\vec{k}_3$.

When these momentum configurations are put in as external momenta
for the loop contributions in Fig.\ \ref{5dias}, they impose 
further restrictions on the loop variables as well because of the 
scale restrictions on the propagators.  
In some of the diagrams, the internal loop variable remains free to explore the 
full shell around the FS. These diagrams will give the leading contributions to the 
flow in the limit $\Lambda/ E_F \to 0$. They are the {\em a)} particle-particle diagrams 
with zero total wavevector with dominant contribution for zero total incoming 
frequency, therefore we keep $V_\Lambda (k,-k,k')$;  and {\em b)} the particle-hole 
diagrams with zero wavevector transfer, where the largest contribution (which is 
possibly the only nonzero one) arises for zero frequency transfer. Hence we keep 
the forward scattering processes $V_\Lambda (k,k',k)$ and $V_\Lambda (k,k',k')$.
The renormalization of all other processes not belonging to one of the three classes 
have less low energy phase space and will be suppressed by a factor $\Lambda/ E_F$. This can be formalized nicely in a $1/N$-expansion\cite{feldman,shankar,chen}.  

We emphasize that an additional argument is needed when using this reasoning 
for the 1PI scheme because the full propagator, which appears on some 
of the lines in the diagrams, is not restricted to scale $\Lambda$, 
but has support on all $k$ with $|e(\vec k)|$ above $\Lambda$. 
It is the decay of the propagator as one moves away from the Fermi surface
that singles out the contribution of the above diagrams, 
where both internal momenta are close to scale $\Lambda$, 
as leading.

The steps of reducing the full RG equations to the ones giving 
Eliashberg theory are as follows. 

\begin{itemize}
\item The Hartree terms for the self-energy (diagram c) in Fig. \ref{5dias}) will only 
give a frequency-independent contribution which might reshape the Fermi surface. 
As we are interested in isotropic systems, the only effect is a 
constant shift which can be absorbed in a redefinition of $\mu$. 
Thus, for calculating  dynamical effects like the $Z$-factor, we may drop the 
Hartree term. Then the self-energy is only of Fock-type (diagram c) in Fig. \ref{5dias}), 
as in Fig. \ref{EliashFig}. 

\item The particle-hole terms for the flow of the interactions contain two diagrams 
which correspond to vertex corrections (see diagrams f) in Fig. \ref{5dias}) 
are subleading. These will be dropped. For an interaction vertex
corresponding to the exchange of acoustic phonons, we can invoke Midgal's theorem, 
which states that in that case, vertex corrections are of order
$c/v_F$, where $c$ is the velocity of sound. Note that in this special situation, 
no small--$\Lambda$ argument is needed since this holds at 
all scales. The mixed contributions, those from other phonon types 
and those due to the electron-electron interactions can be argued to be absorbed 
already in effective vertices \cite{scalapino}, or, more 
rigorously, be shown to generate only subleading terms in the 
particle--particle flow by overlapping loop estimates \cite{FST}.

\item The bubble particle-hole diagram (diagram g) in Fig. \ref{5dias}) corresponds 
to screening of the Coulomb force or to a renormalization of the phonon propagator. 
These effects will already be included in the realistic phonon spectrum which is 
normally used in Eliashberg calculations. Hence we drop these terms as well.
\end{itemize}
Hence the remaining diagrams are the Fock term, the particle-particle diagram and 
the crossed particle-hole ladder (diagrams b), d) and e) in Fig \ref{5dias}). 
A closer look at the them reveals that 
the direct forward scattering is renormalized by the vertex corrections f) and 
bubble diagram g) which we have dropped already. Hence, in this approximation, 
the direct forward scattering does not flow in the normal state. This consistent 
with also neglecting the Hartree diagram for the selfenergy, which would contain 
the direct forward scattering.
The exchange forward scattering $V_\Lambda (k,k',k')$ flows with the crossed 
particle-hole diagram e), which we have kept. Although the contributions in this channel 
are generally small for $\Lambda \gg T$, a rapid flow can develop\cite{metzner} for 
$\Lambda  \sim T$. The exchange forward scattering is the interaction occurring in 
the Fock diagram for the normal selfenergy, which we keep as well.

A strong flow can also develop in the Cooper channel with zero total incoming 
wavevector and frequency, possible leading to a Cooper instability. The corresponding 
processes $V_\Lambda (k,-k,k')$  are renormalized by the particle-particle diagram c) 
in Fig. \ref{5dias}. 
The particle-particle ladder with zero total momentum renormalizes the interaction 
that appears in the anomalous self energy. It is responsible for the growth of the 
superconducting gap amplitude at the Cooper instability.

Both types of interactions, the Cooper pair scattering $(k,-k) \to (k',-k')$ and the 
exchange forward scattering $(k,k') \to (k',k)$, are special in the sense that 
they are determined by only two wavevectors/frequencies instead of three. The 
exchange forward scattering is described by the vertex functions 
$\gaf^{(4)}_\Lambda (K,FK,K',FK')$  where the first two entries belong to 
one incoming (e.g., $K=(ik_0,\vec{k},s,-)$) and one outgoing particle 
(e.g., $FK=(ik_0,\vec{k},s',+) $). $FK$ differs from $K$ by the Nambu index 
$\pm$ and possibly by the spin index $s'$. For pairing between $K$ and $PK$ 
($PK$ reverts frequency and wavevector of $K$, but leaves the Nambu index unchanged, 
the spin $Ps$ depends on the type of the pairing, for singlet pairing it is reversed) 
the Cooper channel is described by $\gaf^{(4)}_\Lambda (K,PK,K',PK')$. If 
we require spin-rotation invariance, the vertices $\gaf^{(4)}_\Lambda (K,FK,K',FK')$ 
and $\gaf^{(4)}_\Lambda (K,PK,K',PK')$, which do not conserve the total spin, are zero. 

Denoting the collision partner $FK$ or $PK$ generally by $\tilde K$, we see that the 
flow due to the two ladder diagrams can be expressed in one ladder-type equation for 
the forward+Cooper processes,
\begin{equation} 
\dot{\gaf}^{(4)}_\Lambda (K,\tilde K,K',\tilde K') = - \frac{1}{2} \sum_{K''} 
 \gaf^{(4)}_\Lambda (K,\tilde K,K'',\tilde K'') 
\frac{d}{d\Lambda} \left[ {\bf G}_\Lambda(K'', \tilde K'') {\bf G}_\Lambda (\tilde K'', K'') \right] \, \gaf^{(4)}_\Lambda (K'',\tilde K'',K',\tilde K') \, . 
\label{oneladder}
\end{equation}
This means that we drop the two other contributions in Eq. \ref{gamma4dot}. 
In those terms $K$ and $\tilde K$ would occur each in a different vertex 
of the two $\gaf^{(4)}$ on the right hand side, hence restricting the low energy 
phase space on the internal lines. 

\begin{figure}
\begin{center}
\includegraphics[width=.5\textwidth]{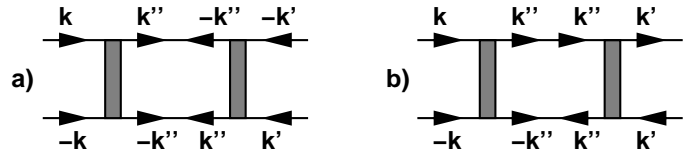}
\end{center}
\caption{Anomalous interaction vertices generated by anomalous Gorkov propagators and forward+Cooper scattering processes. In a) a 4:0 vertex with four incoming lines is generated by two anomalous propagators on the internal lines, in b) one normal and one anomalous propagator create a 3:1 vertex with 3 incoming and one outgoing line. The new vertices fall into the same category of forward+Cooper scatterings as described by Eq. \ref{oneladder}.}
\label{anomal}
\end{figure}

The nice feature of Eq. (\ref{oneladder}) is that it not only describes the flow of 
the normal 2:2 interaction vertices with two incoming and two outgoing lines. 
It also captures the flow of anomalous vertices which are generated when we include 
an initial gap amplitude $\Delta_{\Lambda_0} (k)$ into the flow which breaks the 
global U(1) invariance explicitly,
and which is sent to zero after the flow is performed, to induce
spontaneous symmetry breaking \cite{gapflow}. Furthermore, even if 
spin-rotation invariance is assumed in the normal state, Eq. \ref{oneladder} can be used to 
describe situations where the total spin is not conserved, e.g. when the superconducting 
state prefers a certain spin direction.
In any case, the new vertices generated by Eq. \ref{oneladder} still belong to the 
class of forward and Cooper scatterings and can be described by pairs of generalized 
wavevectors $K$ and $\tilde K$. 
Therefore Eq. \ref{oneladder} remains a closed set and the diagrams dropped above are not generated in the flow.
For nonzero $\Delta_{\Lambda_0} (k)$, the internal propagators also have anomalous 
(Gorkov) contributions which are offdiagonal in Nambu space. Then it is easy to see 
that ladder diagrams with two anomalous propagators (see Fig. \ref{anomal} a)) 
in Eq. \ref{oneladder} generate also $4:0$ or $0:4$ vertices with four incoming 
or outgoing lines. 
The $4:0$ vertices turn out to be essential\cite{gapflow} to stop the flow of the 
gap amplitude for $\Lambda \to 0$. Moreover diagrams with one normal and one anomalous 
propagator (see Fig. \ref{anomal} b)) lead to $3:1$ and $1:3$ vertices with only one 
incoming or outgoing line. For certain situations these vertices can be argued to be 
absent\cite{gapflow} or small, but in general it will be interesting to explore the 
consequences of these anomalous terms. Later we will see that for the special case 
of the ladder flow, the $4:0$ and $3:1$ vertices, although nonzero, disappear again 
out of the final gap equations. This explains why these vertices are usually not 
encountered in standard BCS-type theories.   

The flow of the selfenergy can be expressed with the same class of generalized forward and Cooper vertices,
\begin{equation} 
\dot {\bf \Sigma}_\Lambda (K,\tilde K) = - \frac{1}{2} \sum_{K'} 
 \gaf^{(4)}_\Lambda (K,\tilde K,K',\tilde K') \, {\bf S}_\Lambda(K', \tilde K')
 \, . 
\label{oneself}
\end{equation}
Here we have assumed that translational invariance on the lattice and in imaginary time is not broken. Therefore only normal Nambu-diagonal selfenergies $\Sigma_{11} (ks,ks')$ and $\Sigma_{22} (ks,ks')$, and the anomalous Nambu-offdiagonal selfenergies $\Sigma_{21} (ks,-ks')$ or $\Sigma_{12} (ks,-ks')$ are nonzero. Eqs. \ref{oneladder} and \ref{oneself} reduce to simple BCS model of Ref. \onlinecite{gapflow} if we set the forward scattering to zero and simplify the effective pairing interaction to a constant $V_{\mathrm{eff}}(k_1,k_2,k_3)=-g<0$.  

Thanks to its ladder structure, Eq. \ref{oneladder} can be solved for any scale $\Lambda$ by the Bethe-Salpeter--like equation
\begin{eqnarray}
\gaf^{(4)}_\Lambda (K,\tilde K,K',\tilde K') 
&=& 
\gaf^{(4)}_{\Lambda_0} (K,\tilde K,K',\tilde K') \nonumber 
\\
&& - \gaf^{(4)}_{\Lambda_0} (K,\tilde K,K',\tilde K')
\left[ \frac{1}{2} {\bf G}_\Lambda(K'', \tilde K'') {\bf G}_\Lambda (\tilde K'', K'') \right] \, \gaf^{(4)}_\Lambda (K'',\tilde K'',K',\tilde K') \, . 
\label{oneBS}
\end{eqnarray}
Now we insert this solution into Eq. \ref{oneself}. This gives
\begin{eqnarray} 
\dot {\bf \Sigma}_\Lambda (K,\tilde K) & =&  -\frac{1}{2} \sum_{K'} 
 \gaf^{(4)}_{\Lambda_0} (K,\tilde K,K',\tilde K') \, {\bf S}_\Lambda(K', \tilde K') \nonumber \\
&&  + \frac{1}{2} \sum_{K',K''}
 \gaf^{(4)}_{\Lambda_0} (K,\tilde K,K'',\tilde K'') \left[ \frac{1}{2} {\bf G}_\Lambda(K'', \tilde K'') {\bf G}_\Lambda (\tilde K'', K'') \right] \, \gaf^{(4)}_\Lambda (K'',\tilde K'',K',\tilde K') {\bf S}_\Lambda(K', \tilde K')
\nonumber \\
&=&  - \frac{1}{2} \sum_{K'}
 \gaf^{(4)}_{\Lambda_0} (K,\tilde K,K',\tilde K') \left\{ 
{\bf S}_\Lambda(K', \tilde K') + {\bf G}_\Lambda(K', \tilde K') 
\dot {\bf \Sigma}_\Lambda (K',\tilde K') {\bf G}_\Lambda (\tilde K', K') \right\} 
\nonumber \\
&=&   - \frac{1}{2} \sum_{K'}
 \gaf^{(4)}_{\Lambda_0} (K,\tilde K,K',\tilde K') \,
\dot {\bf G}_\Lambda (K', \tilde K') 
 \, . 
\label{dlamself}
\end{eqnarray}
In going from the first to the second equation, we have inserted Eq. \ref{oneself}. For the last line, we have used Eq. \ref{gdots} and translational invariance. We can immediately integrate this equation from $\Lambda_0$ down to $\Lambda=0$ and obtain
\begin{equation}
{\bf \Sigma}_{\Lambda=0} (K,\tilde K)-  {\bf \Sigma}_{\Lambda_0} (K,\tilde K)  = 
- \frac{1}{2} \sum_{K'} \gaf^{(4)}_{\Lambda_0} (K,\tilde K,K',\tilde K') \, {\bf G}_{\Lambda=0} (K', \tilde K')  \, .  \label{oneeliash} 
\end{equation}
Now we can let ${\bf \Sigma}_{\Lambda_0} (K,\tilde K) \to 0$.
Resolving the Nambu indices we arrive at two familiar equations. For the normal selfenergy $\Sigma (k,s)= \Sigma_{11,\Lambda=0} (K,\tilde K)$ for $K = \vec{k}, ik_0, s,+$ and  $\tilde K = \vec{k}, ik_0, s',-$  we get, picking the Fock contribution (and again assuming spin-singlet pairing for simplicity),
\begin{equation}
\Sigma  (k)  = 
-  \sum_{k'} V_{\Lambda_0} (k,k',k') \, G (k')  \, .  
\label{sigeq} 
\end{equation}
The factor $1/2$ goes away in the summation over the internal Nambu indices.
Defining the offdiagonal propagator $F(k) = G_{21} (k,s-k,-s)$ and the gap function  $\Delta (k) = \Sigma_{21} (k,s,-k,-s)$ we also get the second selfconsistency equation 
\begin{equation}
\Delta  (k)  = 
-  \sum_{k'} V_{\Lambda_0} (k,-k,k') \, F (k')  \, .  
\label{gapeq} 
\end{equation}
These equations are equivalent to Eqs. \ref{E1}  and \ref{E2}. 
Therefore the Eliashberg theory is contained in the RG approach when the 
flow is restricted to the ladder-type diagrams in Cooper and exchange forward scattering.
The anomalous vertices with unequal numbers of incoming and outgoing lines have 
disappeared again from the equations. 
For establishing the connection to the selfconsistent equation we did not have 
to calculate them at all. The equivalence of ladder-type RG equations to selfconsistent 
equations was shown for the normal state by Katanin\cite{katanin}, who pointed 
out that in this case the RG flow fulfills the respective Ward identities,
and for the superconducting state in Ref.\ \onlinecite{gapflow}.
In our case the Ward identity for the global U(1) invariance holds and the 
emerging Goldstone boson can be identified in the flow of the interactions\cite{gapflow}.

Note that after confining the analysis to the separate channel 
for forward and Cooper scattering, respectively, no further assumptions have 
gone into the derivation. In particular, no simplification of the 
frequency or wavevector dependences are necessary. Furthermore, the approximation of 
the normal selfenergy in terms of a $Z$-factor is not essential to establish the 
correspondence of the renormalization group and the self-consistent Eliashberg formalism. 
We also note that the ladder equation (\ref{oneladder}) and the selfenergy equation 
(\ref{oneself}) already go beyond Eliashberg theory.  Eq. \ref{oneladder} also contains 
the bubble summation for the 
direct forward scattering $(ks,k's') \to (ks,k's')$ of the diagrams g) in Fig. \ref{5dias}, 
and the corresponding Hartree selfenergy,  which is diagram c) in Fig. \ref{5dias}. 
These terms are usually not considered in Eliashberg theory, for the reasons indicated 
above. 
We have left out these terms in the RG as well, in order to 
exhibit the approximations involved in obtaining 
Eliashberg theory most clearly. 
Including them is analogous to the procedure for the exchange forward scattering 
and easy because it just adds a Hartree term to Eq. \ref{sigeq}.

Let us briefly discuss the wavevector and frequency dependences and how they develop 
during the flow from $\Lambda_0$ to $\Lambda=0$. For this we assume that the initial 
pairing interaction is factorizable in the incoming and outgoing indices,
\begin{equation} 
V^c_{\Lambda_0} (k,k') = \sum_\ell V_{\ell,\Lambda_0} \,  g_\ell (k) g_\ell(k') 
\end{equation}
with orthonormal functions $g_\ell(k)$ in wavevector-frequency space. Then it is not 
difficult to see that with the assumed ladder structure of the flow in the Cooper channel, 
the RG equations decouple into separate equations for the coefficients $V_{\ell,\Lambda}$. 
Positive $V_{\ell, \Lambda}$ will decrease during the flow. 
This also includes the $s$-wave repulsion due to the 
Coulomb interaction as described by Morel and Anderson\cite{AM}. 
It will be interesting to see if additional insights about this process can be learned 
with the RG approach. 
Negative $V_{\ell,\Lambda}$ will increase until a Cooper instability is reached the 
most negative $\ell$-channel at a critical scale $\Lambda^c_\ell$. 
At this point the normal pairing interaction $V_\Lambda(k,-k,k')$ becomes large. 
Above this scale, if we start with a small initial gap amplitude $\Delta_0$, i.e. small 
off-diagonal propagators, the additional anomalous couplings and the gap amplitude 
$\Delta_\Lambda (k)$ will have departed only little from their initial values. 
However at $\Lambda^c_\ell$ they develop a strong flow due to the almost-divergence 
of $V_\Lambda (k,k')$. Now it is clear from Eqs. \ref{oneladder} and \ref{oneself} 
that the $k$-dependence of the anomalous couplings and $\Delta_\Lambda (k)$ follows 
the $k$-dependence of the divergent component in the $l$-channel, given by $g_l$. 
In fact, if we start with an initial gap 
$\Delta_0 (k)= \sum_\ell \Delta_{0,\ell} g_\ell (k)$, 
only the $l$-component with a Cooper instability at this scale will be pulled 
up and converge to a nonzero value for $\Lambda \to 0$ in the limit $\Delta_0(k) \to 0$.  
All other $\ell$-components in the final gap amplitude $\Delta_{\Lambda=0} (k)$ 
without a Cooper instability in their channel  disappear for $\Delta_0(k) \to 0$. 
The divergence of  the interactions is stopped just below $\Lambda^c_\ell$ by the 
rapid growth of the gap. The final value of the normal interaction vertex scales 
$\propto 1/\Delta_{0,\ell}$. In practice this bound can be used to keep the couplings 
near the perturbative range. The convergence of the results in the limit 
$\Delta_0(k) \to 0$ has been discussed in Ref. \onlinecite{gapflow}. 
All other $\ell$-components in 
the final gap amplitude $\Delta_{\Lambda=0} (k)$ without a Cooper instability 
in their channel  disappear for $\Delta_0(k) \to 0$. 
The resulting gap function corresponds exactly to the one obtained by solving the 
Eliashberg gap equation (\ref{gapeq}) with the most attractive component $V_\ell$ only.
It will be interesting to see how subsequent transitions due to subdominant 
pairing channels are described in this scheme.

\section{Discussion} \label{sec5}
We have shown that a suitable truncation of the 
fermionic functional renormalization group serves as a basis
for the Eliashberg equations both in the normal and symmetry--broken phase, 
thus extending the result of Ref. \onlinecite{tsai} to temperatures
below the critical temperature. 
In our view our rederivation of the Eliashberg equations 
from an approximation to the RG is of methodical interest ---
we have described how, by dropping terms, the RG equations reduce
to a system that integrates to give the self--consistency equations.
Namely, 
integrating the flow of the interactions and selfenergies in the ladder approximation 
described above from the initial bandwidth $\Lambda_0$ down to $\Lambda = 0$ produces 
the same answer as solving the selfconsistent equations (\ref{E1}) and (\ref{E2}) which 
the Eliashberg theory is based on. 
For many cases the latter procedure will still be easier because one 
can work with bare (effective) interactions and does not have to keep track of running 
coupling functions. 
The valuable advantage  of the fermionic functional RG approach presented here 
and in Ref. \onlinecite{tsai} is, however, that it paves the way for the inclusion of 
a number of effects beyond Eliashberg theory.  
This has not been done yet, but while more work is underway let us briefly outline 
some straightforward extensions. 

Note that the Eliashberg equations (\ref{sigeq}) and (\ref{gapeq}) could be derived 
only under the assumption of ladder-like effective vertices in the forward and Cooper 
scattering channel. In contrast with that, the flow equations (\ref{sigmadot}) and 
(\ref{gamma4dot}) hold more generally. 
The difference becomes particularly clear when we consider situations 
where the initial 
interaction does not contain any attractive component in the pairing channel. 
Then 
Eliashberg theory and the corresponding ladder approximation to the flow 
will not find a superconducting solution. 
However, it is known since the work of Kohn 
and Luttinger\cite{kolu} that particle-hole corrections 
to the pair scattering will 
always create an attractive component in the effective interaction. 
In our case, in Eq. \ref{oneladder}, the only 
coupling of the particle-hole processes into the pair scattering is due to the overlap 
between forward and Cooper scattering which gets small due to phase space restrictions.  
This last argument applies only at very low scales $\Lambda$, 
and it is in neglecting the attractive effective interaction generated by 
the integration of fields at higher scales that these approximations 
miss out the Kohn-Luttinger effect.
The more careful analysis of the RG flow, which takes the 
higher scales (regime 1 in Ref. \onlinecite{salmhofer}) into account correctly,
includes the perturbative corrections to the pair 
scattering, in principle to all orders in the initial interaction,
hence renders a 
correct picture of the existence and relative strength of pairing instabilities 
that are dynamically generated.  
For example, in the two-dimensional Hubbard model on the square lattice near 
half-filling it is known that particle-hole corrections generate an attractive 
interaction in the $d_{x^2-y^2}$-wave channel\cite{dwave},
and this becomes especially clear in RG studies
\cite{zanchi,halboth,honerkamp}.
The extension of the flow beyond the Cooper instability opens a way to analyze 
the gap structure of the $d$-wave pairing in more detail. 

In addition to the detection of pairing channels that are not present in the 
initial interaction, the full one-loop flow allows one to study the influence of 
vertex corrections due to the electron-electron interactions on the pairing in a 
systematic way.
For example, the particle-hole channel renormalizes the $s$-wave pairing channel 
as well. This effect, basically due to the same diagrams as the Kohn-Luttinger effect, 
is known to have quantitative consequences even for small interactions\cite{gorkov}. 
Furthermore Migdal's theorem, although proven for acoustic phonons, 
has been argued to break down under various other circumstances\cite{grimaldi}.

We emphasize that the concepts described above are not limited to symmetry breaking 
in the superconducting channel. Other types of flows into long-range ordered states 
can be performed as well. In the context of low-dimensional electron systems, 
the analysis of the interference between magnetic and superconducting order seems 
another promising route.

Finally we address the interesting issue which types of runaway flows can be 
brought to a safe end by the inclusion of some symmetry breaking field in the initial conditions; this also indicates some limitations of this method. 
By adding a static offdiagonal part to the selfenergy one assumes that
that there is long--range order in the system, and this is a rather strong
assumption. In an exact treatment, the behaviour of correlations 
at various scales has to be
described by a time-- and space--dependent order parameter field, 
while the 
above ansatz corresponds to a field independent of space and time. 
In low--dimensional systems, and  under very general conditions, 
long range order gets destroyed by long--wavelength fluctuations 
of the order parameter fields, so that
long range order gets replaced by a Kosterlitz--Thouless like phase for
two--dimensional superconductors or phases with even faster decay of the 
correlations of the order parameter field, if they are described by 
nonlinear sigma models at low energies. 
In certain one-dimensional models like Hubbard 
ladder systems\cite{lin}, the RG often flows to strong coupling 
but all correlations remain short ranged even for $T\to 0$.
The absence of long--range correlations in this case is due to a 
combination of the above--mentioned effects and the competition of
different interactions. One may expect the above ansatz of a static 
symmetry--breaking component to extend to a regime where the dynamics 
of the order parameter fields takes place on much larger temporal
and spatial scales than that of the fermions, so that one can hope to 
describe some aspects of phases without long range order. 
For the understanding of the vicinity of the transition, 
and the situation where strong fluctuations persist
to very low scales, one needs to generalize this scheme by 
including the order parameter fields themselves in the description. 

{\em Acknowledgments:} We thank P. Horsch, A. Katanin, W. Metzner and R. Zeyher 
for useful discussions. The Erwin-Schr\"odinger Institute (ESI) in Vienna is 
acknowledged for hospitality.

\end{document}